\documentclass[%
 reprint, nofootinbib,
 amsmath,amssymb,
 aps,
]{revtex4-1}

\usepackage{scrextend}
\usepackage{subfigure}
\usepackage[applemac]{inputenc}
\usepackage{graphicx}
\usepackage{dcolumn}
\usepackage{bm}

\graphicspath{{imagenes/}}
\usepackage{physics}

\begin{document}


\title{Dirac equation in exotic spacetimes}

\author{Javier Faba Garc\'ia}
\affiliation{Departamento de F\'isica Te\'orica, Universidad Complutense de Madrid, Plaza de Ciencias, 1 28040 Madrid, Spain
}%
\author{Carlos Sab\'in}
\affiliation{Instituto de F\'isica Fundamental, CSIC,
Serrano, 113-bis,
28006 Madrid, Spain}\date{\today}

\begin{abstract}

We find solutions of the Dirac equation in curved spacetime. In particular, we consider 1+1 dimensional sections of several exotic metrics: the Alcubierre metric, which describes a scenario that allows faster-than-light (FTL) velocity; the G\"odel metric, that describes a universe containing closed timelike curves (CTC); and the Kerr metric, which corresponds to the metric of a rotating pointlike source, for instance a rotating black hole. Moreover, we also show that the techniques that we use in these cases can be extended to nonstatic metrics.

\end{abstract}

\pacs{Valid PACS appear here}
\maketitle

\section{Introduction}

In 1928, Paul Dirac formulated a special relativistic version of the wave equation in quantum mechanics ($\hbar=1, c=1$):

\begin{equation*}
i\gamma^\mu\partial_\mu\psi-m\psi=0,
\end{equation*}
where, in the standard representation,
\begin{equation*}
\begin{split}
&\gamma^0 = 
\begin{pmatrix}
1 & 0 & 0 & 0 \\
0 & 1 & 0 & 0 \\
0 & 0 & -1 & 0 \\
0 & 0 & 0 & -1
\end{pmatrix}
\qquad
\gamma^1 =
\begin{pmatrix}
0 & 0 & 0 & 1 \\
0 & 0 & 1 & 0 \\
0 & -1 & 0 & 0 \\
-1 & 0 & 0 & 0
\end{pmatrix}
\\
&\gamma^2 = 
\begin{pmatrix}
0 & 0 & 0 & -i \\
0 & 0 & i & 0 \\
0 & i & 0 & 0 \\
-i & 0 & 0 & 0
\end{pmatrix}
\qquad
\gamma^3 =
\begin{pmatrix}
0 & 0 & 1 & 0 \\
0 & 0 & 0 & -1 \\
-1 & 0 & 0 & 0 \\
0 & 1 & 0 & 0
\end{pmatrix}.
\end{split}
\end{equation*}

This equation was born as an attempt of linearization of the Klein-Gordon equation, that comes from the application of the Einstein relation $E^2=p^2+m^2$ to the Schr\"odinger equation Hamiltonian. The Dirac equation was a turning point in physics. Joining Quantum Mechanics and Relativity, this equation predicts the spin of the electron, the existence of antimatter etc. and it is considered as the natural transition between relativistic quantum mechanics and quantum field theory \cite{thallerdir}. Moreover, a renewed interest in the Dirac equation has come from the field of quantum simulations, which has enabled the experimental observation of some of its more interesting features in tabletop experiments \cite{gerritsma, lamata,klein1,klein2,potwopot,salger2011klein,zitterbecbegung}.
Nevertheless, even though the general solutions of this equation are well-known, this is not the case of the Dirac equation in curved spacetime. In general, obtaining solutions in curved space is more complicated. For a detailed analysis of the 3+1 Dirac equation in Riemannian spacetimes, see for instance \cite{ref1}. Some particular 1+1 D cases can be found, for instance, in \cite{angelakis} and \cite{sabin}.

In this work, the method explained by one of us in \cite{sabin} will be applied to obtain solutions of the Dirac equation in curved spacetime for different exotic metrics in 1+1 dimensional sections, with the help of known solutions in Minkowski spacetime. If we take a 1+1 dimensional section of the spacetime, the Dirac equations reduces to \cite{angelakis}:

\begin{equation*}
i\bigg(\partial_t+\frac{\dot{\Omega}}{2\Omega}\bigg)\psi=-i\sigma_x\bigg(\partial_x+\frac{\Omega'}{2\Omega}\bigg)\psi+\sigma_z\Omega m\psi
\end{equation*}
where $\Omega$ is the conformal factor -- note that any metric in 1+1 dimensions allow a coordinate change such that it acquires the form $ds^2=\Omega^2(-dt^2+dx^2)$ -- the dot and prime stand for time and spatial derivatives, respectively and $\sigma_x$ and $\sigma_z$ are the Pauli matrices:

\begin{equation*}
\sigma_x = 
\begin{pmatrix}
0 & 1 \\
1 & 0
\end{pmatrix}
\qquad
\sigma_z = 
\begin{pmatrix}
1 & 0 \\
0 & -1
\end{pmatrix}
\end{equation*}

Note that if $\Omega= 1$, we obtain the well-known Dirac equation in 1+1 dimensional flat spacetime (see, for instance \cite{ref2}). This equation has been studied in detail, including striking properties such as Zitterbewegung or Klein paradox \cite{ref3,ref4}. Given $\Omega$, we could try to find solutions of this partial differential equation system. However, that procedure could be very involved, depending on the particular form of $\Omega$. The method presented in \cite{sabin} allows to find the solution of this partial differential equation system through a coordinate change and an existing solution in Minkowski spacetime. Thus, the method will be applied to several exotic metrics of interest, such as the Alcubierre metric, which allows FTL velocity; to the G\"odel metric, which allows CTC; and the Kerr metric, which describes the spacetime geometry generated by a rotating body (rotating black holes). Furthermore, we will see that, although in \cite{sabin} $\dot{\Omega}=0$ (static spacetime) is an apparent requirement for the validity of the method, in fact the method is also valid for nonstatic spacetimes.

Our results might be of interest in the context of quantum simulation of the Dirac equation. Several ideas have been proposed for simulating the Dirac equation in Minkowski spacetime for 1+1 dimensional sections: for example, experiments have been already realized with trapped ions \cite{gerritsma,klein2} and Bose-Einstein condensates \cite{salger2011klein,zitterbecbegung}, while realistic proposals exist with superconducting circuits \cite{julen} . Also, some ideas have been proposed for simulating the Dirac equation in curved spacetime for 1+1 dimensional sections: for example, through a mapping between this equation and a multiphoton quantum Rabi model \cite{pedernales} . However, since we obtain solutions in curved spacetime through transformations of Minkowski solutions, the results obtained in this work could be helpful for simulations of Dirac equation in non trivial spacetimes with currently existing setups.

The structure of the paper is the following. We start in Section \ref{sec:Metodo} by recalling the procedure for obtaining solutions of the Dirac equation in curved spacetimes out of flat-spacetime solutions, first introduced in \cite{sabin}. Then, we proceed to apply it in Section \ref{sec:results} to the aforementioned curved spacetimes of interest. Finally, we introduce a generalization of the method to nonstatic spacetimes in Section \ref{sec:nonstatic}, which might be useful for further applications. We conclude in Section \ref{sec:conclus} with a summary of our results. 

\section{Method}
 \label{sec:Metodo}
 We summarize here the procedure to obtain solutions of the Dirac equation in curved spacetimes, first discussed in \cite{sabin}.
 In general, the Dirac equation in curved spacetime is obtained by replacing the standard partial derivatives in the Minkowski equation by the corresponding covariant ones, where the affine connection would carry the dependence on the particular form of the metric. In 1+1 D, using the conformally-flat form of the metric and with a little algebra (see the details in \cite{angelakis}), we can get:

\begin{equation}
\label{eq:DIRAC_MASLESS}
i\bigg(\partial_t+\frac{\dot{\Omega}}{2\Omega}\bigg)\psi=-i\sigma_x\bigg(\partial_x+\frac{\Omega'}{2\Omega}\bigg)\psi.
\end{equation}
For the static case, in which the conformal factor is time-independent $\dot{\Omega}=0$, the equation is:
\begin{equation}
\label{eq:Dirac}
i\partial_t\psi=-i\sigma_x\bigg(\partial_x+\frac{\Omega'}{2\Omega}\bigg)\psi.
\end{equation}
If we make the transformation $\psi = \Omega^{-1/2}\phi$, the equation (\ref{eq:Dirac}) becomes:
\begin{equation*}
i\partial_t\phi=-i\sigma_x\partial_x\phi
\end{equation*}
which corresponds to the Dirac equation for a massless particle in Minkowski spacetime, whose solution is well-known (see, for instance, \cite{thaller}). Thus, using this transformation and the conformal factor of the 1+1 dimensional metric, we can find analytical solutions for the Dirac equation in curved spacetime. So, for a given metric, we have to follow this three-step procedure:

1) Find a change of coordinates $(t,x)\rightarrow (\bar{t},\bar{x})$ such that, in the new coordinates, the metric acquires the form $ds^2=\Omega^2(-d\bar{t}^2+d\bar{x}^2)$.

2) Using those new coordinates $(\bar{t},\bar{x})$, apply $\psi(\bar{t},\bar{x}) = \Omega^{-1/2}(\bar{x})\phi(\bar{t},\bar{x})$, where $\phi(\bar{t},\bar{x})$ is the solution of the Minkowski spacetime Dirac equation.

3) Once having $\psi(\bar{t},\bar{x})$, apply the coordinate change $(\bar{t},\bar{x})\rightarrow(t,x)$ to finally find the solution of the curved spacetime Dirac equation in $(t,x)$ coordinates.

Please note that, in spite of its apparent simplicity, in general it might not be necessarily straightforwerd to perform the steps 1) and 3). To write the metric in a conformally-flat form, it might be needed to express the change of coordinates in a differential form, which could generate an equation that might still not be straightforward to solve, as we will see in the (\ref{sec:BH}) section.

This method was applied in \cite{sabin} to find solutions of the Dirac equation in a 1+1 dimensional section of a traversable wormhole spacetime. Let us now consider other examples of interest.

\section{Results}
\label{sec:results}
\subsection{G\"odel and Alcubierre metric}

Both metrics will be discussed in the same section because of their similarity.

In 1994, Alcubierre proposed a metric that, in principle, allows FTL motion \cite{alcubierre}. Note that, in general relativity, FTL speed is only forbidden locally. This is not as exotic as it might seem at first glance: for instance, the expansion of the Universe can make that two distant galaxies move at FTL speed between them, while each one is moving locally inside its light-cone. While the oposite might be  possible too: if the spacetime were contracting fast enough, each galaxy were moving near the speed of light locally (inside its light-cone) in oposite directions, but globally both were getting closer. With these considerations in mind, Alcubierre's idea is simple: to create, in the front of an object, a spacetime contraction, and in the back, a spacetime dilation. Thus, the contraction will pull the object forward, and the dilation will push the object forward too. Locally the object will be inside its light-cone, but due to this spacetime manipulation, it would move FTL-- as compared with $c$, the speed of light in flat-spacetime vacuum.

The Alcubierre metric inside the ``bubble" created by the space-temporal contraction/dilation, under the limit $\sigma\rightarrow \infty$ \cite{alcubierre}, and taking a spatial section $y=y_0$ and $z=z_0$ (with constant $y_0, z_0$), acquires the form \cite{alcubierre,sabin2}:

\begin{equation}
\label{eq:Alcubierre}
ds^2 = -(1-v_s^2)dt^2+dx^2-2v_sdxdt.
\end{equation}

As expected, Eq. (\ref{eq:Alcubierre}) becomes the Minkowski metric when $v_s=0$. Now we make a coordinate change $(t,x)\rightarrow(\bar{t},\bar{x})$, such that:

\begin{equation*}
\begin{split}
dt&=d\bar{t} \\
dx&=d\bar{x}+v_sd\bar{t}.
\end{split}
\end{equation*}

The Alcubierre metric, using those new coordinates, acquires the form $ds^2=-d\bar{t}^2+d\bar{x}^2$, that is simply the Minkowski metric, so the conformal factor will be $\Omega^2=1$, and the solution of the curved spacetime Dirac equation, $\psi(\bar{t},\bar{x})$, will be equal to the Minkowski spacetime solution, $\phi(\bar{t},\bar{x})$. Now, due to the big similarity -- regarding to the application of our techniques -- between the Alcubierre and G\"odel metric, we first discuss the latter case and then we compare both results. It is relevant to mention that, although the Dirac equation in a G\"odel universe has already been considered in the study of tachyon's stability \cite{konoplya}, our objective is to find explicit analytical solutions for this equation in the G\"odel's background, using our procedure.

K. G\"odel, in 1949, found a solution of the Einstein equations \cite{Godel} corresponding to an homogeneous mass distribution that rotates at each point of the space \cite{kajari}. That distribution of matter causes unusual effects, such as the existence of CTC. In cylindrical coordinates \cite{kajari}, the metric is given by the following expression: 

\begin{equation}
\label{eq:Godel_METRIC}
\begin{split}
ds^2 = &dt^2 - \frac{dr^2}{1+(\frac{r}{2a})^2}-r^2\bigg(1-\Big(\frac{r}{2a}\Big)^2\bigg)d\phi^2-dz^2 \\
&+\frac{2r^2}{a\sqrt{2}}dtd\phi \\
\end{split}
\end{equation}
where $a$ is a parameter with units of length, that represents a characteristic distance. In particular, $r_G=2a$ represents the critical radius from which CTC can exist \cite{kajari}.

Now, taking a radial section with $\phi = \phi_0$ and $z=z_0$ \cite{sabin2}, the G\"odel metric becomes:

\begin{equation*}
ds^2 = dt^2-\frac{1}{1+\big(\frac{r}{2a}\big)^2}dr^2
\end{equation*}
By making the coordinate change $(t,r)\rightarrow(\bar{t},\bar{r})$, with
\begin{equation}
\label{eq:Godel_coordinates}
\begin{split}
d\bar{r}^2&=\frac{1}{1+\big(\frac{r}{2a}\big)^2}dr^2 \\
d\bar{t}^2&=dt^2
\end{split}
\end{equation}
the previous metric transforms into the Minkowski metric $ds^2=d\bar{t}^2-d\bar{r}^2$. Furthermore, we can find the relationship between the $(t,r)$ and $(\bar{t},\bar{r})$ coordinates by performing the following integration:

\begin{equation*}
\bar{r}(r) = \int \frac{dr}{\sqrt{1+\big(\frac{r}{2a}\big)^2}}
\end{equation*}
whose solution -- setting $\bar{r}(r=0)=0$-- is:
\begin{equation*}
\bar{r}(r) = 2a\sinh^{-1}\big(\frac{r}{2a}\big).
\end{equation*}

With these new coordinates, the conformal factor becomes $\Omega^2=1$, and the Minkowski spacetime solution will be the same as the curved spacetime solution, as in the case of the Alcubierre metric. Thus, the only difference between G\"odel and Alcubierre metric is the relationship between the original coordinates and the new coordinates. 

\begin{figure}
\centering
\includegraphics[width=0.5\textwidth]{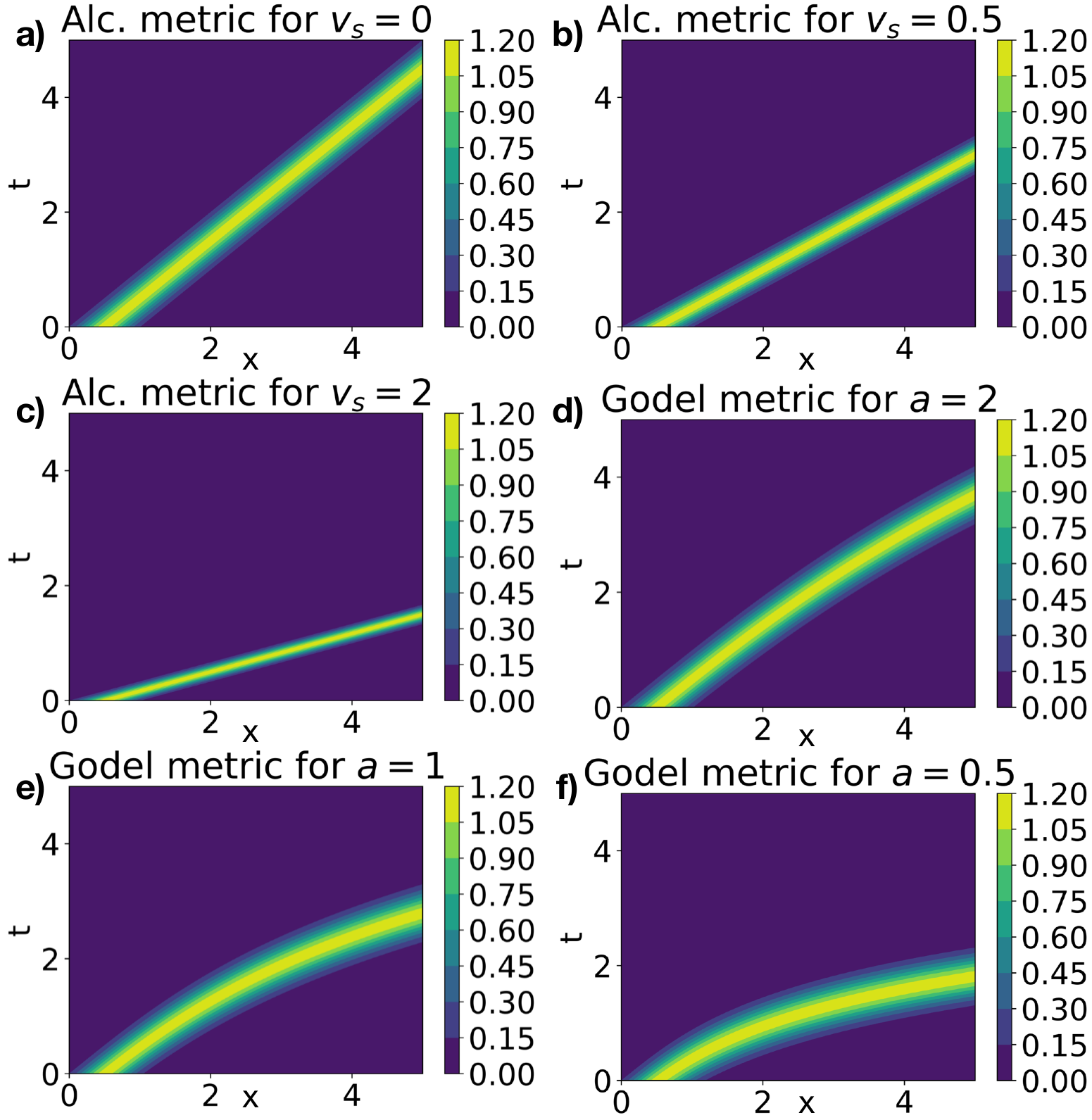}
\caption{spacetime diagrams for the probability density of Dirac equation solutions in a-c) Alcubierre metric with a) $v_s=0$ (equivalent to Minkowski spacetime), b) $v_s=0.5$ (subluminal spacetime bubble), c) $v_s=2$ (superluminal spacetime bubble); d-f) G\"odel metric with d) $a=2$, e) $a=1$, f) $a=0.5$. Please note that the wave function is not normalized.} \label{fig:ALCUBIERRE_Godel}
\end{figure}

Now, if we assume that the wave function has a gaussian initial form:

\begin{equation*}
\phi(\bar{x},0) = Ne^{-\frac{(\bar{x}-\bar{x}_0)^2}{\sigma^2}}
\begin{pmatrix}
1 \\
1
\end{pmatrix}
\end{equation*}
where $\bar{x}$ is the conformally flat coordinate and $N$ is a normalization constant, the solution for the Dirac equation in Minkowski's spacetime will be \cite{sabin,thaller}: 

\begin{equation}
\label{eq:FREE_SOLUTION}
\phi(\bar{x},\bar{t}) = Ne^{-\frac{(\bar{t}-(\bar{x}-\bar{x}_0))^2}{\sigma^2}}
\begin{pmatrix}
1 \\
1
\end{pmatrix}
\end{equation}
Since $\psi=\phi$, to find the curved spacetime solution we only have to apply the coordinate change for each metric. In the Alcubierre case we have:

\begin{equation}
\label{eq:ALCUBIERRE_SOLUTION}
\phi\Big(\bar{x}(x),\bar{t}(t)\Big) = Ne^{-\frac{(t-(x-v_st-\bar{x}_0))^2}{\sigma^2}}
\begin{pmatrix}
1 \\
1
\end{pmatrix},
\end{equation}
while, for the G\"odel metric:
\begin{equation}
\label{eq:Godel_SOLUTION}
\phi\Big(\bar{r}(r),\bar{t}(t)\Big) = Ne^{-\frac{(t-(2a\sinh^{-1}(r/2a)-\bar{x}_0))^2}{\sigma^2}}
\begin{pmatrix}
1 \\
1
\end{pmatrix}.
\end{equation}
Both solutions are represented in Figure \ref{fig:ALCUBIERRE_Godel}.

For a), b) and c), as the parameter $v_s$ increases, the slope of the gaussian wavepacket's ``trajectory" decreases. In other words, the particle's velocity increases as $v_s$ is bigger. In fact, if we take a look to the expression (\ref{eq:ALCUBIERRE_SOLUTION}), we can see its equivalence to the solution of the Minkowski spacetime Dirac equation (\ref{eq:FREE_SOLUTION}) if we substitute $t$ with $(1+v_s)t$. This means that, physically, the particle moves with a FTL velocity $\bar{c} = 1+v_s$, as expected for a massless particle since this is precisely the speed of light in this spacetime.

For d), e) and f), we can see that the particle velocity is no longer constant. This variation is inversely proportional to the value of $a$, as expected, since in the limit $a\rightarrow \infty$ the G\"odel metric (\ref{eq:Godel_METRIC}) becomes the Minkowski metric in cylindrical coordinates. In addition, the slope decreases with the propagation of the particle, implying FTL velocity. Of course, in general the existence of CTC --which is the most genuine feature of G\"odel spacetime--  is intimately linked with the possibility of FTL (see, for instance, \cite{mallary}).

\subsection{Rotating black hole}
\label{sec:BH}

The main motivation to study the Kerr metric is the fact that it describes the spacetime geometry of a rotating body, so it can be used to analyse the physics of black holes in a more realistic scenario than the one provided by the Schwarzschild metric. 
Using Boyer-Lindquist coordinates $(t,r,\theta,\phi)$, the Kerr metric is \cite{kerr}\cite{bambi}:

\begin{equation*}
\begin{split}
ds^2 = &-\Big(1-\frac{2Mr}{\Sigma}\Big)dt^2 - \frac{4Mar\sin^2\theta}{\Sigma}dtd\theta + \frac{\Sigma}{\Delta}dr^2 \\
&+ \Sigma d\theta^2 + \Big(r^2+a^2+\frac{2Ma^2r\sin^2\theta}{\Sigma}\Big) \sin^2\theta d\phi^2
\end{split}
\end{equation*}
where
\begin{equation*}
\Sigma (r,\theta) = r^2 + a^2\cos^2\theta \qquad \Delta (r) = r^2+a^2-2Mr
\end{equation*}
If we take a radial section ($\theta = \theta_0$, $\phi = \phi_0$), we obtain \cite{sabin2}:

\begin{equation*}
ds^2 = -\Big(1-\frac{2Mr}{\Sigma (r,\theta_0)}\Big)dt^2  + \frac{\Sigma (r,\theta_0)}{\Delta (r)}dr^2,
\end{equation*}
where $\Sigma (r, \theta_0) $ is a function of the coordinate $r$ (for simplicity, $\Sigma (r, \theta_0) = \Sigma (r) $). So, if we make the coordinate change $ (t, r) \rightarrow (\bar {t}, \bar {x})$ with:

\begin{equation}
\label{eq:BH_coord_cond}
\begin{split}
\frac{\Sigma (r)}{\Delta (r)}dr^2 &= \Big(1-\frac{2Mr}{\Sigma (r)}\Big)d\bar{x}^2 \\
dt^2 &= d\bar{t}^2 \\
\end{split}
\end{equation}
the metric acquires the form
\begin{equation*}
ds^2 = \Omega^2(r)(-d\bar{t}^2 + d\bar{x}^2)
\end{equation*}
with
\begin{equation*}
\Omega^2(r) = 1-\frac{2Mr}{\Sigma (r)}.
\end{equation*}
Due to the conditions (\ref {eq:BH_coord_cond}), $r$ will be a function of $\bar {x} $, whereas $t =\bar {t} $, so the conformal factor $\Omega (r) $ will be a function of $r (\bar {x}) $:

\begin{equation}
\label{eq:BH_omega}
\Omega^2 \big(r(\bar{x})\big) = 1-\frac{2Mr(\bar{x})}{\Sigma \big(r(\bar{x})\big)}
\end{equation}

To find the conformal factor $\Omega \big (r (\bar {x}) \big) $ we need to obtain $r (\bar {x}) $. Using the first equation in (\ref{eq:BH_coord_cond}) and, after doing some algebraic manipulations, we obtain that $\bar {x} (r) $ is given by:

\begin{equation}
\label{eq:BH_integral}
\bar{x}(r) = \int dr\Sigma (r)\sqrt{\frac{1}{\Delta(r)\big(\Sigma (r)-2Mr\big)}} + C
\end{equation}
In general, solving this integral is not straightforward. Later, we will analyze some particular cases in which the integral can be solved. By now, let us just assume that $r (\bar {x}) $ is known. Thus, the conformal factor is be given by the equation (\ref {eq:BH_omega}) and, therefore, the relation between the solution to Dirac equation in curved spacetime ($\psi$) and flat spacetime ($\phi$) will be:

\begin{equation}
\begin{split}
\label{eq:BH_curved_flat_relation}
\abs{\psi (\bar{x},\bar{t})}^2 &= \abs{\Omega^{-1/2}\big(r(\bar{x})\big)}^2\abs{\phi (\bar{x},\bar{t})}^2 \\
&= \abs{\sqrt{\frac{\Sigma \big(r(\bar{x})\big)}{\Sigma \big(r(\bar{x})\big)-2Mr(\bar{x})}}}\abs{\phi (\bar{x},\bar{t})}^2
\end{split}
\end{equation}
All the wave function properties caused by the spacetime curvature are contained in the factor $\Omega ^ {-1} \big (r (\bar {x}) \big) $, so it is interesting to analyze this function. First, there exist regions in which the probability density of the wave function in curved spacetime becomes infinite. It happens when the following condition is fulfilled  -- we relax the notation and write $r$ instead of $r (\bar {x}) $:

\begin{equation}
\label{eq:BH_horizons}
\Sigma (r)-2Mr=0
\end{equation}
This condition is satisfied whenever $r=r_{\pm}$, with:
\begin{equation}\label{eq:ergosphere}
r_{\pm} = M\pm\sqrt{M^2-a^2\cos^2\theta_0},
\end{equation}

which are the well-known apparent singularities where the temporal component of the metric changes sign, defining the ergosphere \cite{gravitation}. On the other hand, there exist points in which the wave function is null, independently of the form of the solution in Minkowski's spacetime. This will happen whenever $\Omega ^ {-1} (r) = 0$, thats is, when:

\begin{equation*}
\Sigma (r) = 0 \qquad \rightarrow \qquad r^2+a^2\cos^2\theta_0 = 0,
\end{equation*}
which will be satisfied when at least one of those conditions apply:

1) $r=0$, $a=0$ (no rotation)

2) $r=0$, $\theta_0 = \frac{\pi}{2}$.

Finally, there is a region of the space in which the conformal factor is an imaginary number. This occurs if:
\begin{equation*}
\Sigma(r)-2Mr<0.
\end{equation*}
This second-degree inequality  is verified for $ r _- <r <r _ + $, namely, within the ergosphere.

\subsubsection{Radial section with $\theta_0 = 0$}
\begin{figure}
\includegraphics[width=\linewidth]{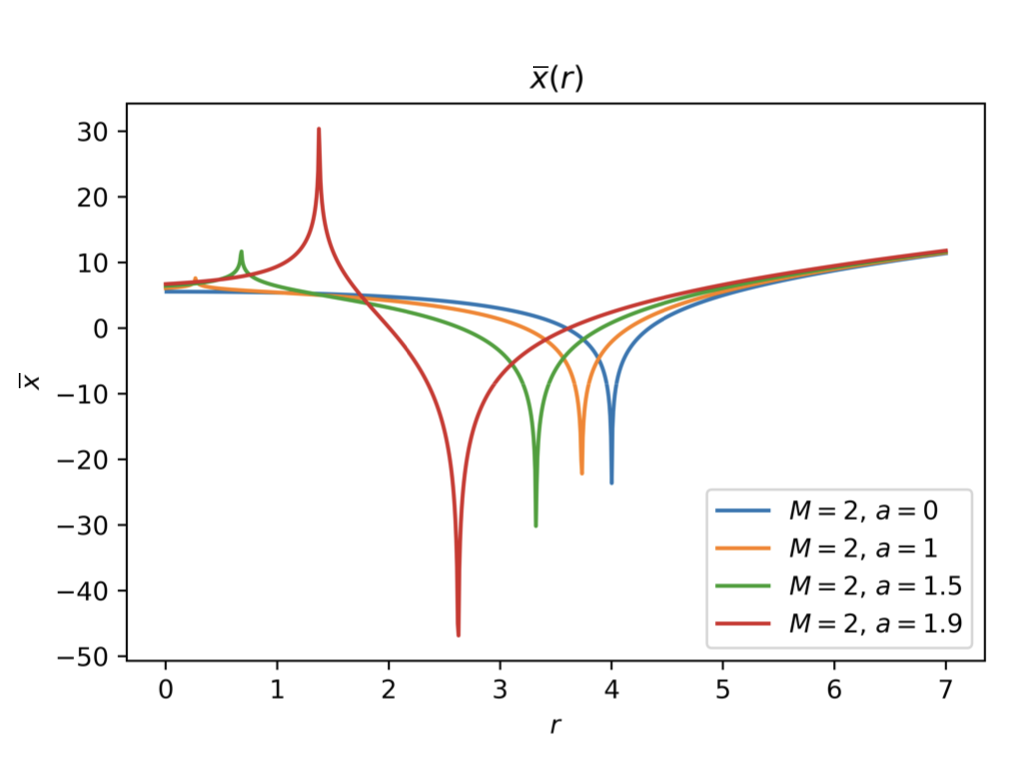}
\caption{$\bar {x} (r) $ for different values of the parameter $a$. In absence of rotation ($a=0$), the singular points are in $r = 0$ and $r = 2M$ (Schwarzschild's black hole). As the rotation increases, both points ($r=r _ {\pm} $) come closer to $r = M$.}
\label{fig:BH_x(r)}
\end{figure}

Let u consider $\theta_0 = 0 $. Then $\Sigma(r) = r^2 + a^2 $, $\Delta(r) = \Sigma-2Mr$, and the integral becomes:

\begin{equation*}
\bar{x}(r) = \int dr\frac{\Sigma(r)}{\Delta(r)}+C=\int dr\frac{r^2+a^2}{r^2+a^2-2Mr}+C
\end{equation*}
whose solution is
\begin{equation*}
\begin{split}
\bar{x}(r) =& \frac{2M^2}{\sqrt{a^2-M^2}}\tan^{-1}\Big(\frac{r-M}{\sqrt{a^2-M^2}}\Big) \\
&+M\log(a^2-2Mr+r^2)+r+C
\end{split}
\end{equation*}
Bearing in mind that, for a rotating black hole, $M^2> a^2$ is satisfied (otherwise, it would not be a black hole, but a highly rotating body \cite{stephani}), and using the relationship $\tan^{-1}(ix) = i\tanh^{-1}(x)$, we can write the solution as:
\begin{equation}
\label{eq:BH_radial_section_solution}
\begin{split}
\bar{x}(r) =& \frac{2M^2}{\sqrt{M^2-a^2}}\tanh^{-1}\Big(\frac{M-r}{\sqrt{M^2-a^2}}\Big) \\
&+M\log(a^2-2Mr+r^2)+r+C
\end{split}
\end{equation}
The constant of integration can be chosen in such a way that the constant imaginary part of Eq. (\ref{eq:BH_radial_section_solution}) cancels out in each of the three spacetime regions defined by the apparent singularities $r_{\pm}$. Thus:
\begin{equation*}
C = \left \{
\begin{split}
-i\frac{\pi M^2}{\sqrt{M^2-a^2}} \qquad &if \qquad r<r_- \\
-iM\pi \qquad &if \qquad r_-<r<r_+ \\
-i\frac{\pi M^2}{\sqrt{M^2-a^2}} \qquad &if \qquad r_+<r \\
\end{split}
\right.
\end{equation*}
In Figure \ref{fig:BH_x(r)}, we plot $\bar{x}(r)$ for different values of $M$ and $a$.  

\begin{figure*}
\centering
\includegraphics[width=\textwidth]{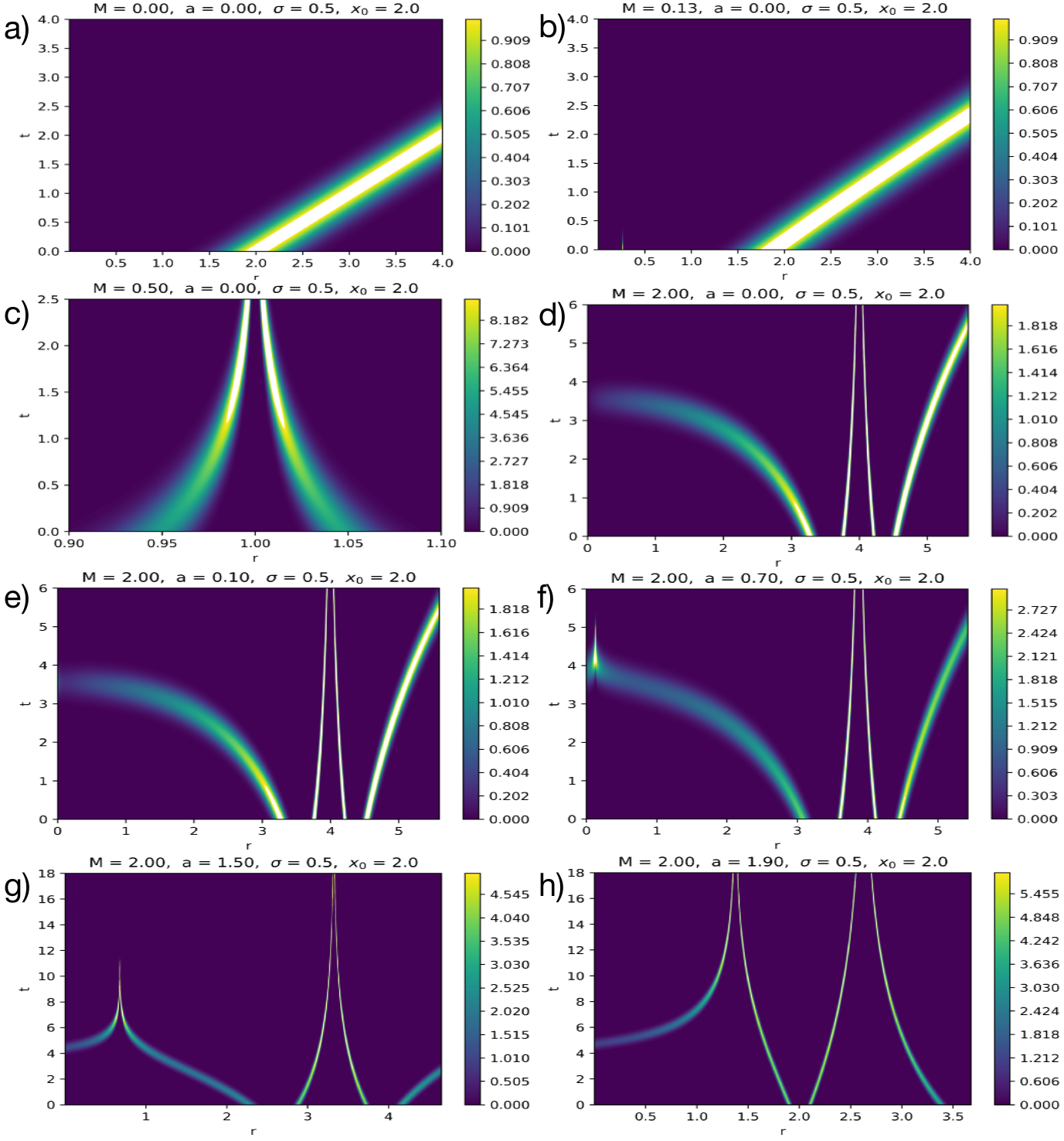}
\caption{Dirac equation solution for Kerr metric and different values of the parameters $a$ and $M$. a)-d), $a=0$ (no rotation, Schwarzschild metric) and $M$ ranging from $0$ to $2$. The horizon is at $r=2M$. e)-h) $M=2$ and $a$ ranging from $0$ to $2$, two singularities appear according to Eq. (\ref{eq:ergosphere}) with $\theta_0=0$ giving rise to an ergosphere which splits the full spacetime into three separate regions \label{fig:BH_solution}. Notice that as $a$ increases $r_+$, $r_-$ move from $2M$ and $0$, respectively, to $M$, where they would merge in the limit $M=a$. Please note that the wave function is not normalized.}
\end{figure*}

%
%
Finally, assuming again that our particle is characterized by a Gaussian wavepacket, the solution in Kerr spacetime is:

\begin{equation*}
\begin{split}
\abs{\psi \big(\bar{x}(r),t\big)}^2 &= \sqrt{\frac{\Sigma (r)}{\Sigma (r)-2Mr}}\abs{\phi \big(\bar{x}(r),t\big)}^2 \\
&=2N^2\sqrt{\frac{\Sigma (r)}{\Sigma (r)-2Mr}}e^{-\frac{2(t-(\bar{x}(r)-\bar{\bar{x}}_0))^2}{\sigma^2}}
\end{split}
\end{equation*}
This solution is represented in Figure (\ref{fig:BH_solution}) for several values of $a$ and $M$.

Naturally, when we set $M=a=0$, the metric becomes the Minkowski one, $\Omega^2 = 1$, and the expression (\ref {eq:BH_radial_section_solution}) reduces to $\bar {x} (r) =r$. In other words, the solution corresponds to a free particle in flat spacetime, as it is observed in Figure (\ref {fig:BH_solution} a). If we slightly increase the value of $M$ (with $a=0$, thus we are in Schwarzschild spacetime), we see that the free solution is slightly disturbed, increasing the probability density near the horizon, as can be seen in Figure  (\ref {fig:BH_solution} b). As the mass increases, the probability density accumulates nearby the horizon, placed in $r=2M$. In addition, we see that the  width of the wavepacket is drastically reduced (the particle is more localized in space), and the particle's speed diminishes, vanishing in the horizon. Finally, if we set the mass to a constant value and we change $a$ (Figures (\ref {fig:BH_solution} e-h), we see that, as $a$ increases, the inner horizon is generated, around which, again, the probability density accumulates in the same way as before. In addition, in the region between both singular points --the ergosphere-- we see how the particle tends to be expelled towards the interior or exterior horizon.

\section{Nonstatic spacetimes}\label{sec:nonstatic}
So far, we have only analyzed static metrics, that is, $\dot {\Omega} =0$. We now show that our techniques are also valid for the case of nonstatic metrics ($\dot {\Omega} \neq0$). Performing the substitution $\psi = \Omega ^ {-1/2} \phi$ in Equation (\ref {eq:DIRAC_MASLESS}) and with a little algebra, we find:
\begin{equation*}
\begin{split}
i\bigg(\partial_t+\frac{\dot{\Omega}}{2\Omega}\bigg)(\Omega^{-1/2}\phi)&=-i\sigma_x\bigg(\partial_x+\frac{\Omega'}{2\Omega}\bigg)(\Omega^{-1/2}\phi) \\
\partial_t(\Omega^{-1/2}\phi)+\frac{\dot{\Omega}}{2\Omega}\Omega^{-1/2}\phi&=-\sigma_x\bigg[\partial_x(\Omega^{-1/2}\phi)+\frac{\Omega '}{2\Omega}\Omega^{-1/2}\phi\bigg]
\end{split}
\end{equation*}
Notice that 
\begin{equation}
\partial_t (\Omega ^ {-1/2} \phi) =-\frac {1} {2} \Omega ^ {-3/2} \dot {\Omega} \phi + \Omega ^ {-1/2} \partial_t\phi.
\end{equation}
Therefore the terms not containing $\partial_t\phi$ cancel out. The same thing happens with the spatial partial derivative: the terms without $\partial_x\phi$ cancel out. Thus, putting everything together we recover the equation:
\begin{equation*}
i\partial_t\phi=-i\sigma_x\partial_x\phi
\end{equation*}
that is the Dirac equation for Minkowski spacetime. Therefore, our techniques could be also applied to nonstatic spacetimes, such as Friedmann-Lemaître-Robertson-Walker spacetime, as in \cite{angelakis}.

\section{Conclusions}
\label{sec:conclus}
Following \cite{sabin}, we have shown that is possible to obtain solutions of the Dirac equation in curved spacetime by means of a transformation
$\psi(\bar{x},\bar{t}) = \Omega^{-1/2}(\bar{x})\phi(\bar{x},\bar{t})$
where $\psi(\bar{x},\bar{t})$ is the curved spacetime solution, $\phi(\bar{x},\bar{t})$ is the Minkowski spacetime solution and $\Omega(\bar{x})$ is the conformal factor. Then, in order to be able to apply this transformation, it is necessary to perform first a coordinate change $(x,t)\rightarrow(\bar{x},\bar{t})$, such that in the new coordinates, the metric is conformally flat.
We have applied this technique to three different metrics: 1+1 dimensional sections of Alcubierre, G\"odel and Kerr metrics. For Alcubierre and G\"odel, the conformal factor turns out to be  $\Omega^2=1$, so the only difference between both solutions is the coordinate transformation employed in each case. Setting a gaussian solution in Minkowski spacetime, and after undoing the coordinate transformation, we obtain the solutions ($\ref{eq:ALCUBIERRE_SOLUTION}$) and ($\ref{eq:Godel_SOLUTION}$), for Alcubierre and G\"odel spacetimes, respectively. In this way, we have been able to analyse the FTL behaviour of massless particles in these exotic spacetimes.
In the case of the Kerr metric, the conformal factor is  non-trivial $\Omega^2(r) = 1-\frac{2Mr}{\Sigma (r)}$. We find an analytical solution in a spatial section with $\theta_0=0$. We have discussed the dependence of this solution on the black-hole mass and rotation, as well as the behaviour of the particle near the singularities and within the ergosphere. Finally, we have shown that our technique is valid also for nonstatic spacetimes.

Our results might, in principle, be useful in the context of quantum simulations of the Dirac equation. In particular, following the spirit of \cite{sabin}, notice that our technique entails that curved spacetimes can be encoded into a transformation realized onto a flat-spacetime Dirac wavepacket. Therefore, in principle, our results could be tested in already existing quantum simulators of the Dirac equation in flat spacetime.

\section*{Acknowledgements}
C. S. has received financial support through the Junior Leader Postdoctoral Fellowship Programme from ``la Caixa" Banking Foundation and Fundaci\'on General CSIC (ComFuturo Programme)

\end{document}